\documentclass[twocolumn]{aastex62}

\submitjournal{ApJ}

\shorttitle{Faraday dispersion function of disk galaxies I}
\shortauthors{Eguchi et al. 2020}

\usepackage{graphicx}
\usepackage{color}
\usepackage{comment}
\usepackage{multirow}
\begin{document}

\title{\huge{Faraday dispersion function of disk galaxies \\ with axisymmetric global magnetic fields I}}

\correspondingauthor{Haruya Eguchi}
\email{schalke1101@gmail.com}

\author[0000-0001-8197-2335]{Haruya Eguchi}
\affil{University of Kumamoto, 2-39-1, Kurokami, Kumamoto 860-8555, Japan}

\author{Masaki Suzuki}
\affil{University of Kumamoto, 2-39-1, Kurokami, Kumamoto 860-8555, Japan}

\author{Yoshimitsu Miyashita}
\affil{University of Kumamoto, 2-39-1, Kurokami, Kumamoto 860-8555, Japan}

\author{Shinsuke Ideguchi}
\affil{Department of Astrophysics/IMAPP, Radboud University Nijmegen, PO Box 9010, NL-6500 GL Nijmegen, the Netherlands}

\author{Keitaro Takahashi}
\affil{University of Kumamoto, 2-39-1, Kurokami, Kumamoto 860-8555, Japan}
\affil{International Research Organization for Advanced Science and Technology, Kumamoto University, Kumamoto 860-8555, Japan}



\begin{abstract}
Faraday tomography is a novel method to probe 3-dimensional structure of magnetic fields of polarized radio sources. In this paper, we investigate intrinsic Faraday dispersion function (FDF) of disk galaxies extending a simple analytic model of galactic magnetic fields developed in \citet{Ideguchi2017}. The model consists of axisymmetric coherent fields and turbulent fields and we consider the effects of inclination, the relative amplitude of coherent and turbulent magnetic fields and pitch angle of coherent fields. Our simple model makes it easy to obtain physical interpretation of FDFs and helps understanding observational results. We find FDFs generally have two peaks when galaxies are observed with non zero inclination and the gap between the two peaks, their relative height and their widths are dependent on the model parameters. Especially, the gap is strongly dependent on the inclination angle and typically varies by a factor of 2 when we consider the inclination of 10 - 60 deg, while the relative height varies only by less than $20\%$. These findings provide us with the important lesson that the presence of two peaks in intrinsic FDFs does not necessarily imply the presence of two separate radio sources within a beam and they allow us to estimate the galactic parameters.
\end{abstract}


\section{Introduction}

Magnetic fields play important roles in many astrophysical systems  including galaxies \citep[e.g.][for a review]{Beck2013,Beck2016,Han2017}. In studies on such objects, 3-dimensional structure of magnetic fields is a key to understand the dynamics of magnetized plasma. Faraday tomography is a technique to reconstruct the distribution of magnetic fields of polarized radio sources along the line of sight (LOS) and a combination with 2-dimensional imaging allows us to probe the 3-dimensional structure \citep{Burn1966,Brentjens2005}. 

A key quantity in Faraday tomography is Faraday depth defined as
\begin{equation}
\phi(x) = K \int_{0}^{x} n_e(x') B_\parallel(x') dx' ~~~ [{\rm rad \ m^{-2}}]
\label{phi}
\end{equation}
where $n_e(x)~[{\rm cm}^{-3}]$, $B_\parallel(x)~[\mu{\rm G}]$ and $x~[{\rm pc}]$ are the thermal electron density and magnetic fields parallel to the LOS and the coordinate along the LOS. The coefficient is $K = 0.81~[{\rm rad \ m^{-2}}]$ when the above units for $n_e$, $B_\parallel$ and $x$ are adopted. Then, the Faraday dispersion function (FDF), the complex polarization intensity in Faraday depth space, is related to the polarization spectrum as,
\begin{equation}
P(\lambda^2) = \frac{1}{\pi} \int_{-\infty}^{\infty} F(\phi) e^{2 i \phi \lambda^2} d\phi.
\label{eq:P}
\end{equation}
The FDF is the quantity which can be obtained from Faraday tomography and contains information on the LOS distribution of magnetic fields, high-energy electrons which emit polarized synchrotron radiation and thermal electrons. Although the distribution which is given by the FDF is in Faraday depth space, rather than physical space, it gives us much more information than the conventional rotation measure \citep{schnitzeler2007,Govoni2010,Mao2010,Wolleben2010,Akahori2014,Anderson2016,O'Sullivan2018}.

However, there are two major difficulties concerning Faraday tomography. The first is the precise reconstruction of the FDF from observed polarization spectrum. Although Eq.(\ref{eq:P}) is mathematically equivalent to Fourier transform, the inverse transformation to obtain the FDF is incomplete because the observational band is limited to a finite range of $\lambda^2 (> 0)$. A lot of algorithms have been developed to extract as much information as possible from observed polarization spectrum: RM CLEAN \citep{Heald2009,Miyashita2016}, QU fitting \citep{O'Sullivan2012,Miyashita2019} and sparse modeling \citep{Li2011,Andrecut2012,Akiyama2019}.
 
The second one, which is directly related to the current work, is the interpretation of the FDF. As we stated above, the FDF is the distribution of polarized emission in Faraday depth space, not physical space. Thus, the interpretation of the FDF is not straightforward, even if it was reconstructed precisely. In the case of galaxies, previous works have emphasized that the presence of turbulent magnetic fields is important to understand the shape of the FDF \citep{Beck2009,Bell2011,Frick2011,Ideguchi2014}. Generally, the turbulent fields induce randomness in the FDF and make the FDF difficult to interpret.

In our previous study \citep{Ideguchi2017}, we developed a simple analytic model of face-on disk galaxies with global (coherent) and turbulent (random) magnetic fields. It was found that randomness of the FDF is significantly reduced when the FDFs of areas which are much larger than the typical scale of turbulence are summed over. The resultant FDF reflects the feature of global fields and the statistical properties of turbulence. While our previous work considered face-on galaxies, it is also important to investigate the effect of inclination of galaxies. In this paper, we extend the model of galaxies developed in \citet{Ideguchi2017} to consider the effects of inclination of observation, relative amplitude between global and turbulent magnetic fields and pitch angle of global fields to FDFs. 

The structure of this paper is as follows. In section 2, we describe our model of galactic magnetic fields and method to calculate FDFs. The results are given in section 3. Section 4 is devoted to the summary and discussion.

\section{Model of disk galaxies}

\subsection{Galaxy Model in \citet{Ideguchi2017}}

In our previous work \citep{Ideguchi2017}, we studied the characteristic properties of the FDF of face-on spiral galaxies. This work followed \cite{Ideguchi2014}, where we used a realistic Galactic model but our interpretation of the FDF was not straightforward because of its complicated behavior mostly due to turbulent magnetic fields. Thus, in \cite{Ideguchi2017}, we employed a toy model of spiral galaxies and made a number of simplifying assumptions to understand the essence of the FDF, rather than to construct a realistic and quantitative model.

In \cite{Ideguchi2017}, we considered a galactic disk with the thickness of $2L_{\rm sh}$\footnote{We referred to the thickness as ``scale height'' in the previous paper and thus, the subscript of the variable was put as ``sh''. However, it should be noted that this does not mean the distribution is exponential but it is actually assumed to be constant as we mention later.}, while we neglected the contribution from the galactic halo or thick disk for simplicity. To calculate the FDF, we need the distribution of the synchrotron emissivity and Faraday rotation measure values which is derived from the thermal electron density ($n_e$) and line-of-sight (LOS) component of magnetic fields ($B_\parallel$). $B_\parallel$ is contributed from random and coherent components such that
\begin{equation}
    B_\parallel = B_{\rm rand} + B_{\rm coh}.
\end{equation}
The model galaxy consisted of fundamental cells of size $L_{\rm cell}$ which corresponds to the typical fluctuation scale of the $B_{\rm rand}$. The above physical quantities were assigned to each cell and were assumed to be uniform for simplicity except for $B_{\rm rand}$ (see below). The cells were stacked up for $[-L_{\rm sh},L_{\rm sh}]$ with the polarized radiative transfer taken into account. We also considered multiple cells ($N_{\perp}\times N_{\perp}$) in the plane of the sky which corresponds to the observational region. As a result, the computational box contains $N_{\perp} \times N_{\perp} \times N$ cells with $N = 2 L_{\rm sh}/L_{\rm cell}$ being the LOS direction. In other words, each LOS includes $N$ ``layers'' which includes $N_{\perp} \times N_{\perp}$ cells. Thus, the size of the computational box is $L_{\rm cell}N_{\perp} \times L_{\rm cell}N_{\perp} \times L_{\rm cell}N$.

As mentioned above, the physical quantities were assumed to be uniform except for $B_{\rm rand}$, and we take the values by referring the past observations. First, for the half thickness of the galactic disk, we took $L_{\rm sh} = 1$ kpc as a representative value. This value is consistent with the scale heights of thermal electron density of the Galaxy \citep{Gaensler2008} and the synchrotron emissivity of external galaxies \citep{Krause2009}.

Regarding the synchrotron polarization emissivity, we assumed that it is constant in the computational box for simplicity, and we did not consider the distributions of cosmic ray electron and plane-of-the-sky magnetic field individually, from which the synchrotron emissivity is calculated. The observations of edge-on galaxies show that the synchrotron emissivity decays exponentially from the galactic disk with the scale height of $\sim$ 1 kpc \citep{Dumke1998,Heesen2009,Irwin2012a,Irwin2013,Wiegert2015,Krause2018} and thus, we took the thickness of the disk ($L_{\rm sh}$) as 1kpc. We discuss the effect of the exponential decay of the emissivity on the results of the current work in Section 4.

For the distribution of the thermal electron density which appears in the calculation of the Faraday rotation, $n_e = 0.02~{\rm cm}^{-3}$ was chosen as a representative value taken from \cite{Gaensler2008}. Although the density is assumed to be uniform in this paper, we discuss the effects of the inhomogeneity on the results in Section 4.

The LOS component of magnetic field ($B_\parallel$) is another factor for the Faraday rotation. We simply set the coherent component as $B_{\rm coh}=0-5~\mu$G and saw its effect on the FDF systematically. The turbulent component, $B_{\rm rand}$, was the only quantity that is not uniform in the model, and random numbers taken from the uniform distribution were allocated. Since the strength of the turbulent magnetic fields is estimated to be $\sim$15 $\mu$G in face-on galaxies \citep{Beck2016}, we set the rms (root-mean-square) strength of $B_{\rm rand}$ to be $15/\sqrt{3}~\mu$G, that is, the rms strength of the three dimensional random field is 15 $\mu$G. Note that we assumed that there is no correlation between the value of $B_{\rm rand}$ of different cells.

In the model, the size of cubic cells $L_{\rm cell}$ corresponds to the typical fluctuation scale of the $B_{\rm rand}$. Observations indirectly imply wide range of $L_{\rm cell}$ values from 10 to 100 pc \citep{Ohno1993,Haverkorn2006,Beck2016}, and we adopted $L_{\rm cell}=10$ pc as a representative value.

Finally, we calculated the FDF from the constructed model and studied the variation by changing the parameters systematically. We found that the FDFs are highly spiky and the different realization of $B_{\rm rand}$ results in a very different shape of the FDF even for the same set of the parameters when $N_\perp^2 \lesssim O(10)$. However, the FDF becomes smoother when $N_\perp^2 \gtrsim 100$ and the physical interpretation becomes much easier. This is because the effects of the randomness of turbulent magnetic fields on the Faraday depth are statistically averaged out. Since we adopted $L_{\rm cell}=10$ pc, this corresponds to $(N_\perp L_{\rm cell})^2$ = (100 pc)$^2$.

As the next step, we derived the FDF analytically in the limit of infinitely large $N_\perp$. We found that the FDF can be expressed as a sum of Gaussian functions with different means and variances:
\begin{equation}
    F(\phi) \propto \sum_{j=1}^N P_j(\phi)
    \label{eq:FDF}
\end{equation}
(Eq. (11) in \citet{Ideguchi2017}), where $P_j(\phi)$ is the $j$th layer's contribution to the FDF and is expressed as
\begin{equation}
    P_j(\phi) = \frac{1}{\sqrt{2\pi j}\sigma_\phi}\exp\left[ -\frac{(\phi-j\Delta\phi_{\rm coh})^2}{2j\sigma_\phi^2} \right]
    \label{eq:layer-org}
\end{equation}
(Eq. (12) in \citet{Ideguchi2017}). Here $\Delta \phi_{\rm coh} = K n_e B_{\rm coh} L_{\rm cell}$ shows the peak location of the emission from the $j$th layer in $\phi$ space which is determined by the coherent field. On the other hand, $\sigma_\phi^2 = K^2 n_e^2 \sigma_B^2 L_{\rm cell}^2$ with $\langle {B_{\rm rand}}^2 \rangle = \sigma_B^2$ shows the variance of the emission from the $j$th layer in $\phi$ space due to the random fields within the layer. Note that we did not consider the absolute emissivity of radiation and thus the intensity of $F(\phi)$ is in arbitrary unit, since we only studied the shape of $F(\phi)$. We also demonstrated that the simulated FDF approaches the analytical solution as the number of LOSs increases.

\subsection{Extension of Galaxy Model}

\begin{figure}[t]
 \includegraphics[width=8cm,clip]{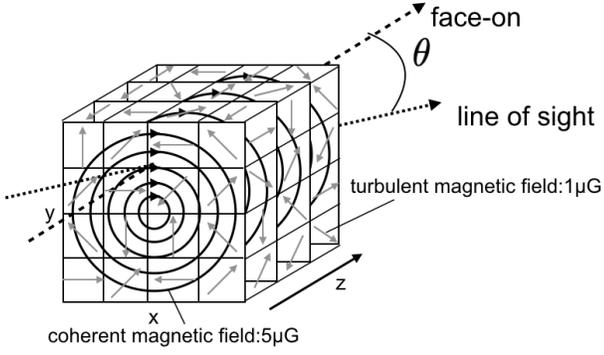}
 \caption{Schematic picture of the galactic model used in this paper. }
 \label{jou}
\end{figure}

\begin{figure}[t]
 \includegraphics[width=8cm,clip]{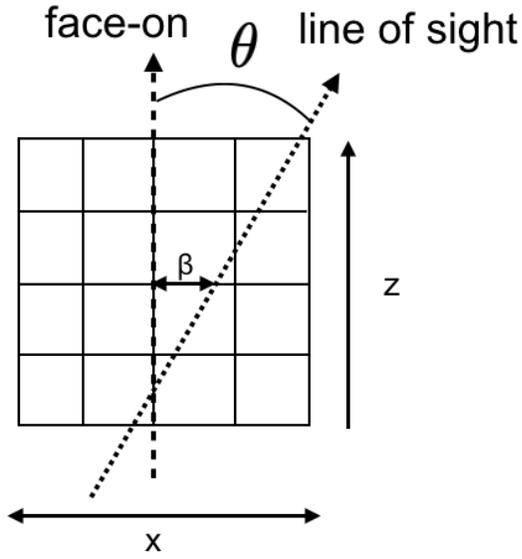}
 \caption{This shows the pattern diagram of our galaxy model viewed from above.}
 \label{model_top}
\end{figure}

In this work, we follow the basic concepts and parameters of the analytical study of our previous work \citep{Ideguchi2017}, but extend it by introducing the inclination of the spiral galaxies as well as the pitch angle of the coherent magnetic fields, and study how these additional ingredients make an impact on the shape of the FDF. We take the values of $n_e = 0.02$ cm$^{-3}$, $L_{\rm sh}=1$ kpc, and $L_{\rm cell}=10$ pc following our previous work. As in Fig.~\ref{jou}, we use Cartesian coordinates $(x,y,z)$, where the $x \mathchar`-y$ plane coincides with the galactic midplane with $z$ penetrating the midplane, and the origin, $(x,y,z)=(0,0,0)$, being put at the galactic center. The inclination angle, $\theta$, represents the angle between the LOS and $z$ axis and the LOS is assumed to be parallel to $x \mathchar`-z$ plane (constant $y$). The location of intersection of the LOS and the galactic plane (at $z=0$) is defined as $(x,y) = (\beta,y)$ (Fig.~ \ref{model_top}). For a fiducial set of parameters, we take $\theta = 40~{\rm deg}$ and $(\beta,y) = (0~{\rm pc},200~{\rm pc})$ but we will see the effect of the variation of these parameters later.

Spiral pattern in the large-scale magnetic field has been observed in many galaxies, not only in spiral galaxies \citep{Braun2010,Beck2016} but also in barred \citep{Beck2012b}, ringed \citep{Chyzy2008} and irregular \citep{Chyzy2000} galaxies. Since the spiral magnetic field can be thought as a ring magnetic field with a non-zero pitch angle, we begin with the simplest case with a ring magnetic field along the galactic plane ($x \mathchar`-y$ plane), and we consider a spiral field by introducing a finite pitch angle later. According to \citet{Fletcher2010}, the mean coherent (regular) magnetic field estimated from 21 nearby galaxies is 5 $\mu$G. So, we set the strength of this ring magnetic field to 5 $\mu$G as a representative value. 
The perpendicular ($B_\perp$) and parallel ($B_\parallel$) components of $B_{\rm coh}$ with respect to the LOS are expressed as
\begin{equation}
    B_\perp(z_j) = \frac{\sqrt{x_j^2 + y^2\cos^2\theta}}{\sqrt{x_j^2+y^2}}B_{\rm coh},
    \label{eq:perp}
\end{equation}
\begin{equation}
    B_\parallel(z_j) = \frac{y\sin\theta}{\sqrt{x_j^2+y^2}}B_{\rm coh},
    \label{eq:parallel}
\end{equation}
where
\begin{equation}
    x_j = z_j\tan\theta+\beta,
    \label{eq:x_j}
\end{equation}
\begin{equation}
    z_j = \left(j-\frac{N}{2}\right)L_{\rm cell}\cos\theta
    \label{eq:z_j}
\end{equation}
(Note that the definition of $B_\parallel$ here is different from that in the previous section: $B_\parallel$ here is the LOS component of $B_{\rm coh}$ while in the previous section, $B_\parallel = B_{\rm rand} + B_{\rm coh}$). Here $(x_j,z_j)$ are the values of $x$ and $z$ of the $j$th layer, $N$ is the total number of layers (see below). Though we do not consider the coherent magnetic field which is perpendicular to the galactic plane, the ring field generates the parallel component with respect to the LOS when $\theta$ is non-zero as can be seen in Eq.~(\ref{eq:parallel}).

Regarding the synchrotron radiation, it should be noted that the emissivity depends on $j$ because the value of $B_\perp$ varies with $j$ according to Eq.~(\ref{eq:perp}). In this paper, we assume that the emissivity of synchrotron radiation is proportional to $B_\perp^{1.8}$. This value of the index is close to the value suggested by the observations of the Milky Way \citep{Reich1988,Remazeilles2015,Planck2016,Krachmalnicoff2018} and external galaxies \citep{Beck1996,Irwin2012b,Klein2018}. 
Although the angle of synchrotron polarization changes along the ring field within the observational region, its variation is not significantly large and it is at most 0.24 radian because we consider the region of (100 pc)$^2$ in the sky plane at the 200 pc (as a representative value) from the galactic center. This variation of polarization angle affects the beam depolarization, but it only changes the amplitude of the FDF and does not change the shape itself.

The random turbulent magnetic field is also assumed. The polarization from its perpendicular component is ignored also in this study, because it is thought to be mostly depolarized due to beam depolarization. Strictly, it is not trivial how completely the beam depolarization occurs since it slightly depends on the direction of coherent field. This small effect will be explored in our future work. Thus, only LOS component is considered for the random fields and is labeled as $B_{\rm rand}$, and it affects the width of emission from each layer in $\phi$ space. \citet{Fletcher2010} estimated from the observation of 21 nearby galaxies that the mean coherent field is 5 $\mu$G as mentioned above and the total field is 17 $\mu$G, which result in $\sim$ 16 $\mu$G for the turbulent field. We start our discussion with a small amplitude of turbulent fields, $B_{\rm rand}=1~\mu$G that gives $\sigma_B=1\mu$G, which make the interpretation of the results relatively easy. Then, later in the next section, we consider larger turbulent fields up to $B_{\rm rand}=10~\mu$G.

Under the assumptions described above, the FDF from $j$th layer can be expressed using Eq. (\ref{eq:layer-org}) as
\begin{equation}
    P_j(\phi) = \frac{B_\perp(z_j)^{1.8}}{\sqrt{2\pi j}\sigma_\phi}\exp\left[ -\frac{(\phi-\phi_0(z_j))^2}{2j\sigma_\phi^2} \right]\times\exp[2\chi(z_j)],
    \label{eq:layer}
\end{equation}
where
\begin{equation}
    \phi_0(z_j) = Kn_eL_{\rm cell}\sum_{k=1}^jB_\parallel(z_k),
    \label{eq:phi0}
\end{equation}
\begin{equation}
    \chi(z_j) = \cos^{-1}\left[\frac{x_j}{\sqrt{x_j^2+y^2\cos\theta}}\right],
    \label{eq:angle}
\end{equation}
and $\sigma_\phi^2=K^2n_e^2\sigma_B^2L_{\rm cell}^2$ with $\langle B_{\rm rand}^2\rangle = \sigma_B^2$. Since we always consider the region for [-$L_{\rm sh}$,$L_{\rm sh}$] through $z$ axis, the number of layers depends on the inclination angle and is expressed as
\begin{equation}
    N = \frac{2L_{\rm sh}}{L_{\rm cell}\cos\theta}.
\end{equation}
Finally, $F(\phi)$ is calculated using Eq. (\ref{eq:FDF}) and (\ref{eq:layer}).

When non-zero pitch angle of spiral field, $\theta'$, is considered, $F(\phi)$ is calculated in the same way by modifying Eq. (\ref{eq:perp}), (\ref{eq:parallel}) and (\ref{eq:angle}) as
\begin{equation}
    B_\perp(z_k) = \frac{\sqrt{x_j^2 + y^2 - (y\cos\theta' + x_j\sin\theta')^2\sin^2 \theta}}{\sqrt{x_j^2+y^2}}B_{\rm coh},
\end{equation}
\begin{equation}
    B_\parallel(z_k) = \frac{(y\cos\theta'+x_j\sin\theta')\sin\theta}{\sqrt{x_j^2+y^2}}B_{\rm coh},
\end{equation}
and
\begin{eqnarray}
&& \chi'(z_k) = \nonumber \\
&& \cos^{-1}\left[\frac{x_j\cos\theta'-y\sin\theta'}{\sqrt{(x_j^2+y^2)\cos^2\theta+(y\sin\theta'-x_j\cos\theta')^2\sin^2\theta}}\right], \nonumber \\
\end{eqnarray}
respectively.

The model parameters and their fiducial values used in this paper are summarized in Table \ref{table:data_type}.

\begin{table*}[t]
  \caption{Fiducial model parameters}
  \label{table:data_type}
  \centering
  \begin{tabular}{lcr}
    \hline
    Symbol & Physical Quantity  &  Fiducial Value  \\
    \hline \hline
    $\theta$  & inclination angle  & 40 deg \\
    $\theta'$  & pitch angle  & 0 deg \\
    y  & height of LOS   & 200 pc \\
    $\beta$  & intersection of LOS and galactic plane & 0 pc \\
    $n_e$  &  thermal electron density & $0.02~{\rm cm}^{-3}$ \\
    $B_{\rm coh}$ & three dimensional coherent magnetic field & $5~\mu {\rm G}$ \\
    $B_{\rm rand}$ & standard deviation of LOS turbulent magnetic fields & $1~\mu {\rm G}$   \\ 
    \hline
  \end{tabular}
  \label{table1}
\end{table*}

\section{Results}

\subsection{basic results}

\begin{figure}[t]
  \includegraphics[width=8cm]{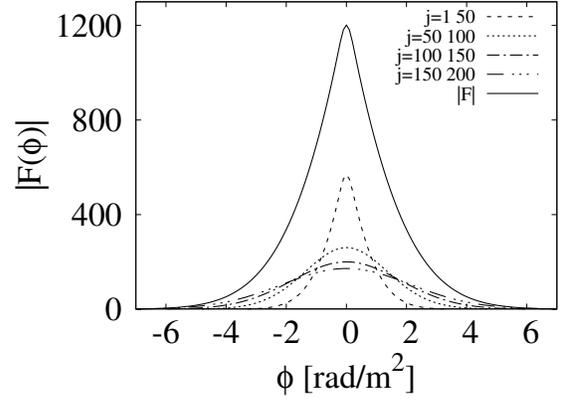}
  \caption{Absolute value of Faraday dispersion function for the case with $\theta = 0~{\rm deg}$ (face-on). Solid line corresponds to the total FDF. Contribution from groups of layers with $j = 1-50,51-100,101-150$ and $151-200$ are also shown. }
  \label{fig:FDF_face-on}
\end{figure}

First, let us see the case of $\theta=0~{\rm deg}$ (face-on) discussed in \citet{Ideguchi2017}. The absolute value of the FDF is shown in Fig. \ref{fig:FDF_face-on}. As explained in Section 2, the FDF is the sum of contribution from all layers along the line of sight. Because ring fields are perpendicular to the line of sight, only turbulent fields contribute to the parallel component of magnetic fields. In this case, the center of Gaussian function of each layer is located at $\phi = 0$, while the width is larger for larger $j$. Thus, the FDF is the sum of Gaussian functions with the same center ($\phi = 0$) and different widths and is not Gaussian itself. The width of the FDF reflects the magnitude of turbulent magnetic fields. \citet{Ideguchi2017} further considered the case with coherent magnetic fields perpendicular to the galactic plane and found that the FDF is skewed. Here, it should be noted that the FDF does not depend on the location in the case of a face-on galaxy.

\begin{figure}[t]
  \includegraphics[width=8cm]{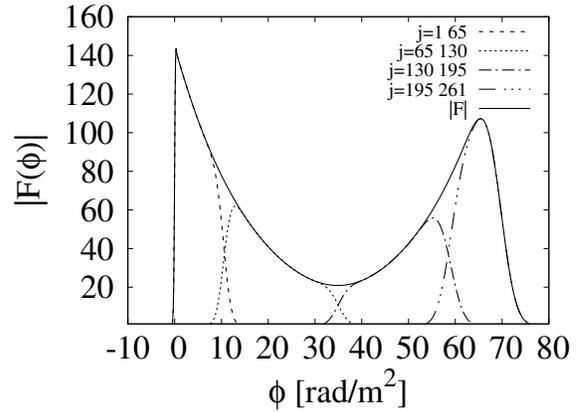}
  \caption{Same as Fig. \ref{fig:FDF_face-on} but for inclination $\theta = 40~{\rm deg}$.}
  \label{fig:FDF_inc40}
\end{figure}

\begin{figure}[t]
  \includegraphics[width=8cm]{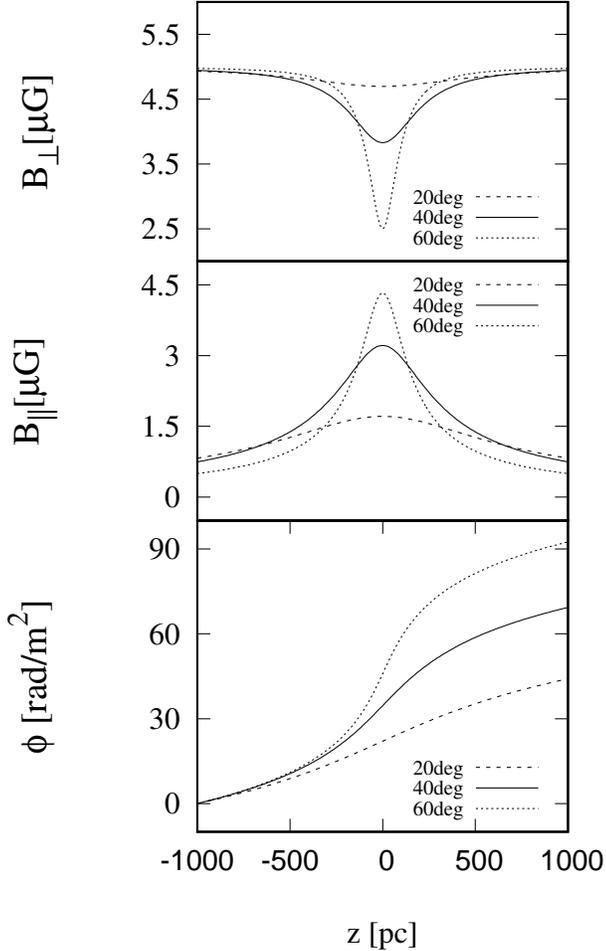}
  \caption{Distribution of $B_\perp$ (top) and $B_\parallel$ (middle), and Faraday depth (bottom) along a line of sight with different inclination $\theta =$ 20, 40 and 60 degree for the fiducial model.}
  \label{fig:distribution}
\end{figure}

Fig. \ref{fig:FDF_inc40} represents the FDF with inclination $\theta = 40~{\rm deg}$ and offset from the galactic center of $y = 200~{\rm pc}$ and $\beta = 0~{\rm pc}$. Two peaks can be seen at $\phi = 0$ and $65~{\rm rad/m^2}$ and the left peak is sharp compared with the other \citep{Beck2012a}. These features were not seen in the situation considered in \citet{Ideguchi2017}.

To understand these features, we plot the distribution of magnetic field components and $\phi(z)$ along the line of sight in Fig. \ref{fig:distribution}. The top panel represents the perpendicular component of coherent fields and it is small around the galactic plane ($z \sim 0$). Because the emissivity of synchrotron radiation is assumed to be proportional to $B_{\perp}^{1.8}$, the emissivity around the galactic plane is consequently relatively low. On the other hand, the distribution of the parallel component is shown in the middle panel and the resulting Faraday depth is plotted in the bottom panel as a function of $z$. Since $\phi(z)$ is monotonically increasing because $B_\parallel > 0$ everywhere, Gaussian functions move rightward with increasing $j$. This can also be seen in Fig. \ref{fig:FDF_inc40}, where the contributions from 4 groups of layers are also shown. Thus, we can see that the peak at $\phi = 0$ is mostly contributed from emission from the near side of the galaxy, while the other is from the far side. The sharpness of the peaks can also be explained by the fact that Gaussian functions of the far side have larger widths.
It should be noted that the dip between the two peaks around $\phi \sim 35$ corresponds to the galactic plane $z \sim 0$ (see the bottom panel of Fig. \ref{fig:distribution}). There are two reasons for the existence of the dip. One is that a relatively few number of layers are contributing to the FDF at $\phi \sim 35$ since $B_\parallel$ is large around $z \sim 0$ and $\phi$ is rapidly increasing as can be seen in the middle and bottom panels of Fig. \ref{fig:distribution}. Another reason is that the emissivity is relatively low there as stated above.

\begin{figure*}[t]
  \includegraphics[width=8cm]{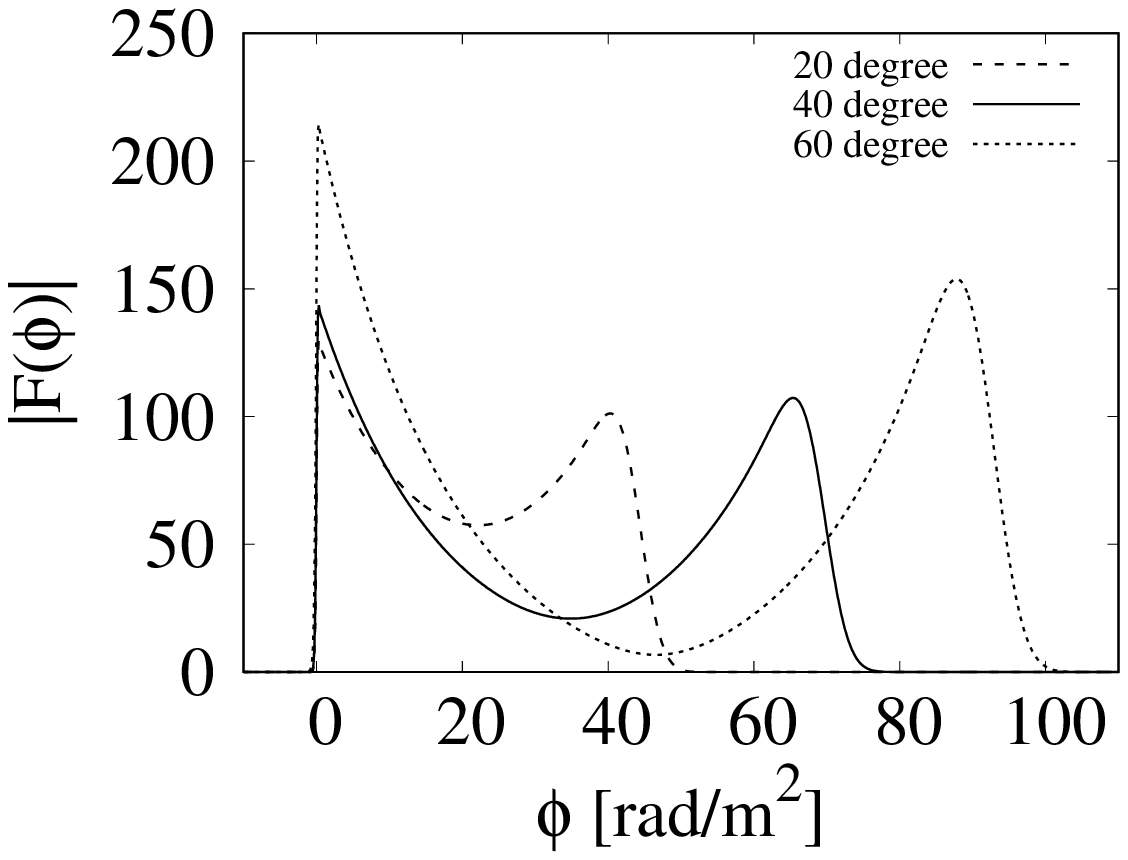}
  \includegraphics[width=8cm]{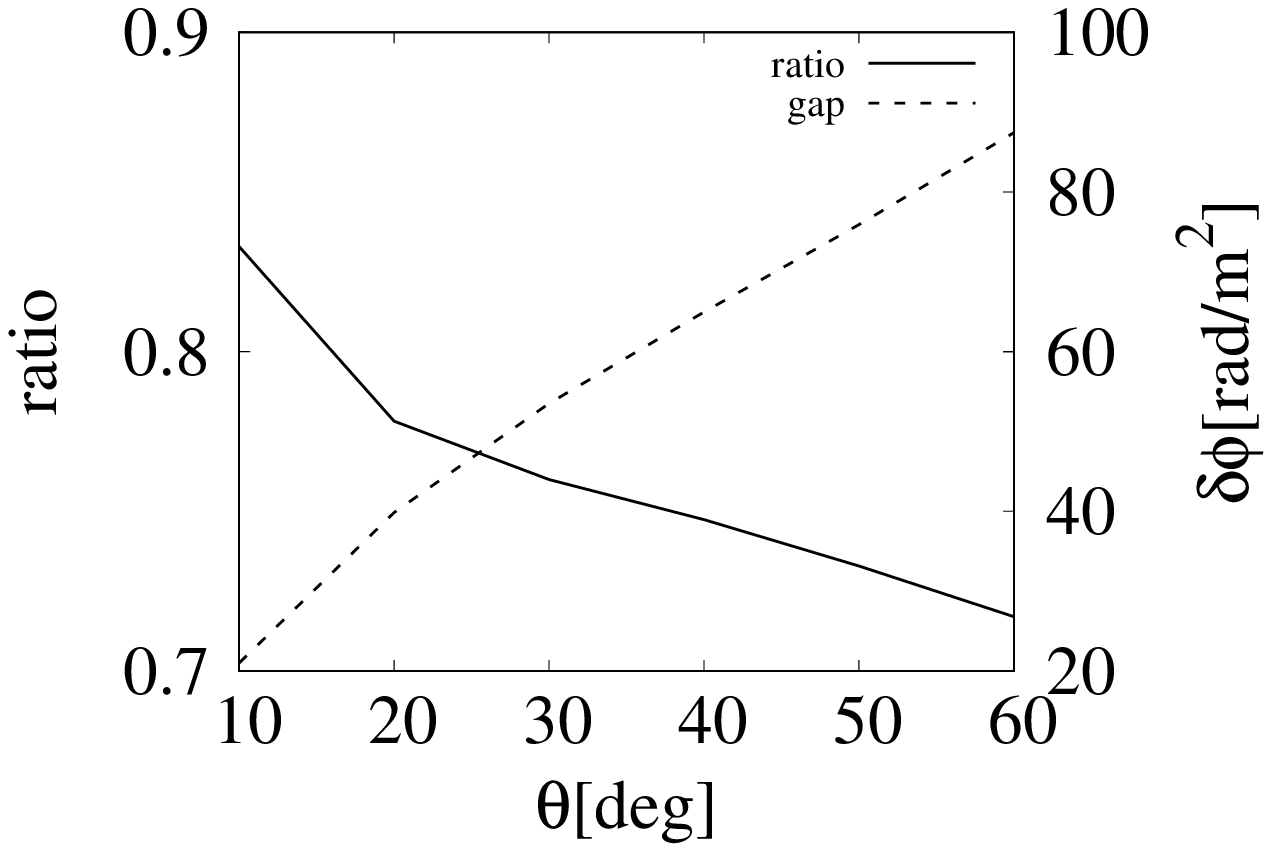}
  \caption{left:Absolute value of Faraday dispersion function for the cases with $\theta =$ 20, 40 and 60 deg. right:The gap between the two peaks and the ratio of the peak heights (right peak/left peak) with varying the inclination $\theta$.}
  \label{fig:FDF_inc}
  \centering
\end{figure*}

The left panel of Fig. \ref{fig:FDF_inc} shows the absolute value of the FDF for different inclination angles ($\theta =$ 20, 40 and 60 deg), and the right panel shows the gap between the two peaks and the ratio of peak heights (right peak/left peak) with varying the inclination $\theta$. From this figure, we see that the gap increases almost linearly with the inclination angle $\theta$ and reaches 90 [rad/m$^2$] when $\theta = 60~{\rm deg}$. This is because the LOS component of magnetic fields around $z \sim 0$ pc becomes larger for a larger inclination by about 1.5 $\mu$G for a change of $\theta$ of 20 deg (see the middle of Fig.~\ref{fig:distribution}). On the other hand, the peak values increase with the inclination angle while the dip becomes deeper, because a smaller number of the layers contributes to the dip for a larger value of $B_{||}$ around $z \sim 0$ pc. Further, the left peak at $\phi \sim 0$ [rad/m$^2$] increases more than the right peak and the peak ratio (right peak/left peak) decreases by about $15\%$ when $\theta$ varies from 20 deg to 60 deg.


\begin{figure*}[t]
  \includegraphics[width=8cm]{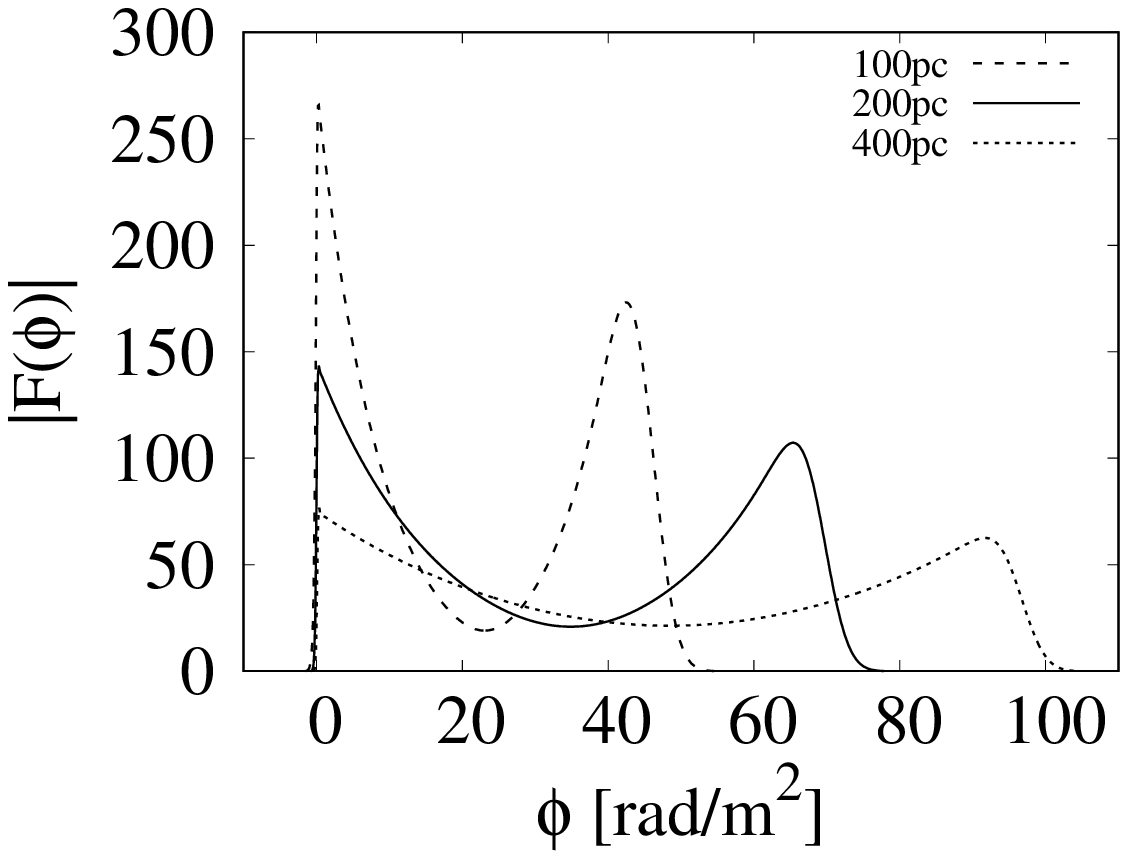}
  \includegraphics[width=8cm]{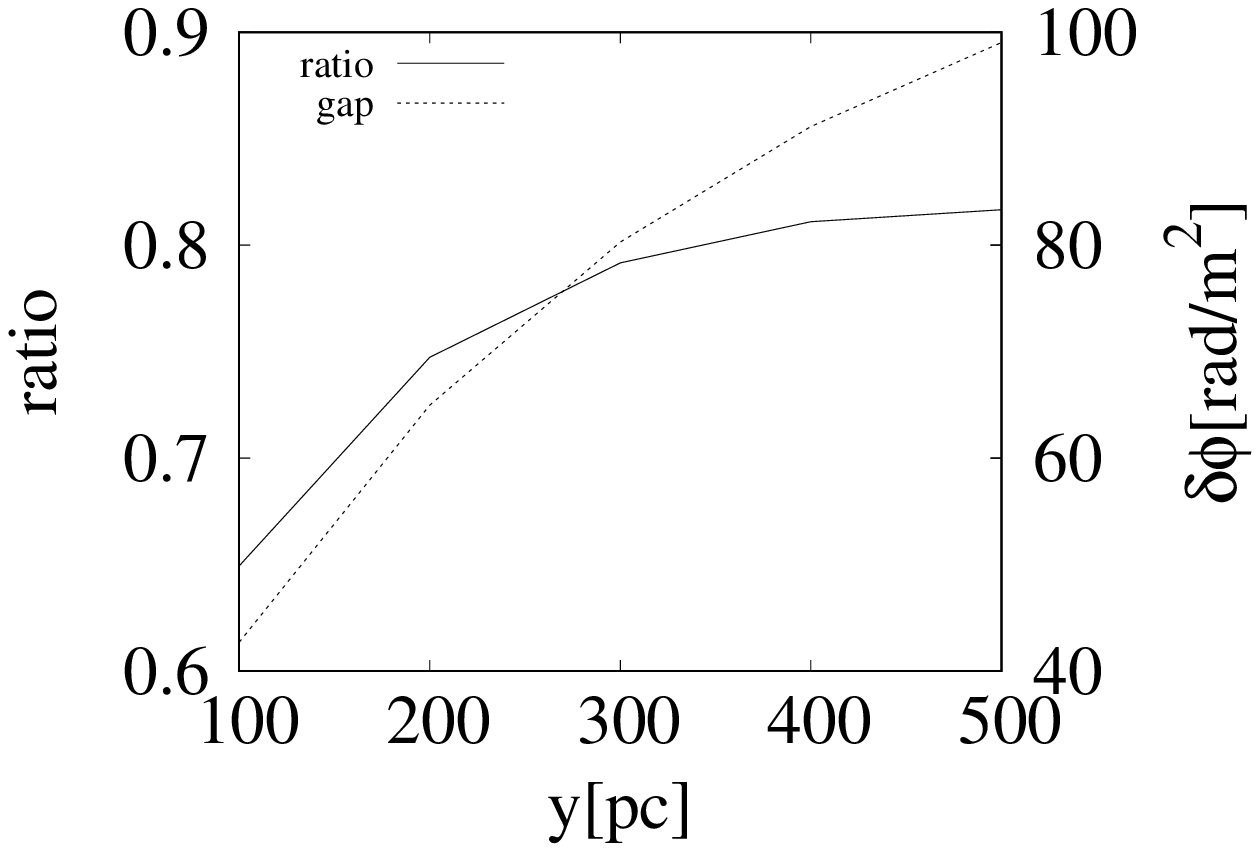}
  \caption{Same as Fig. \ref{fig:FDF_inc} but for varying the offset $y$. Here, we fix the inclination angle $\theta$ to 40~deg and $\beta =$ 0~pc.}
  \label{fig:FDF_offset-y}
\end{figure*}

\begin{figure}[t]
  \includegraphics[width=8cm]{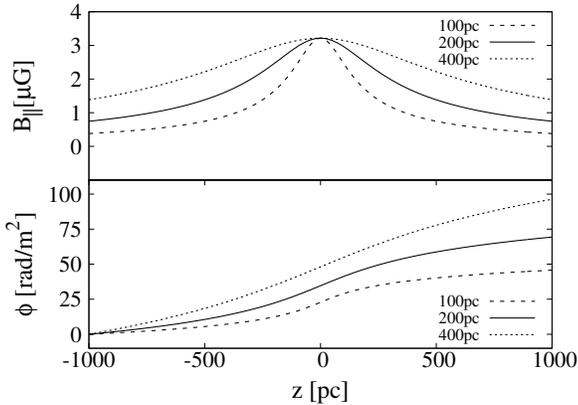}
  \caption{Distribution of $B_\parallel$ (Top), and Faraday depth (Bottom) along a line of sight with different offset $y$ = 100, 200 and 400~pc. The inclination angle $\theta$ is fixed to 40 deg.}
  \label{fig:distribution_offset}
\end{figure}

\begin{figure*}[t]
 \includegraphics[width=8cm]{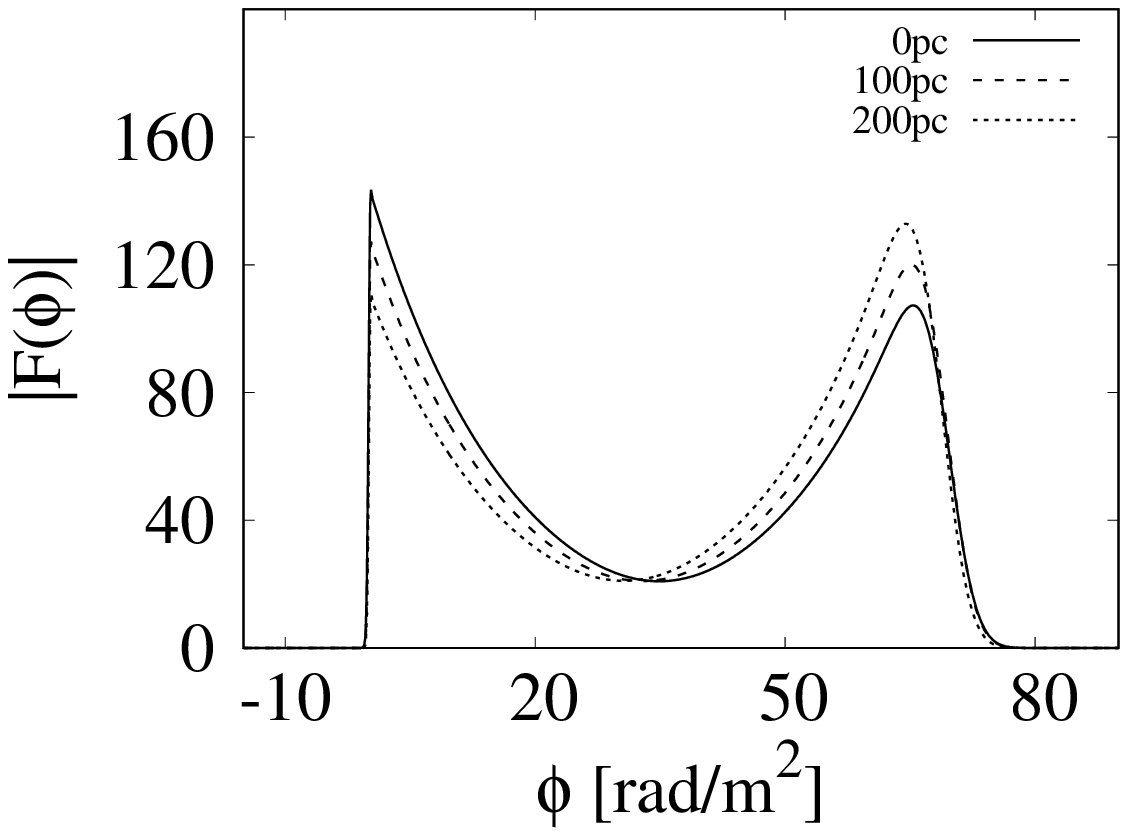}
 \includegraphics[width=8cm]{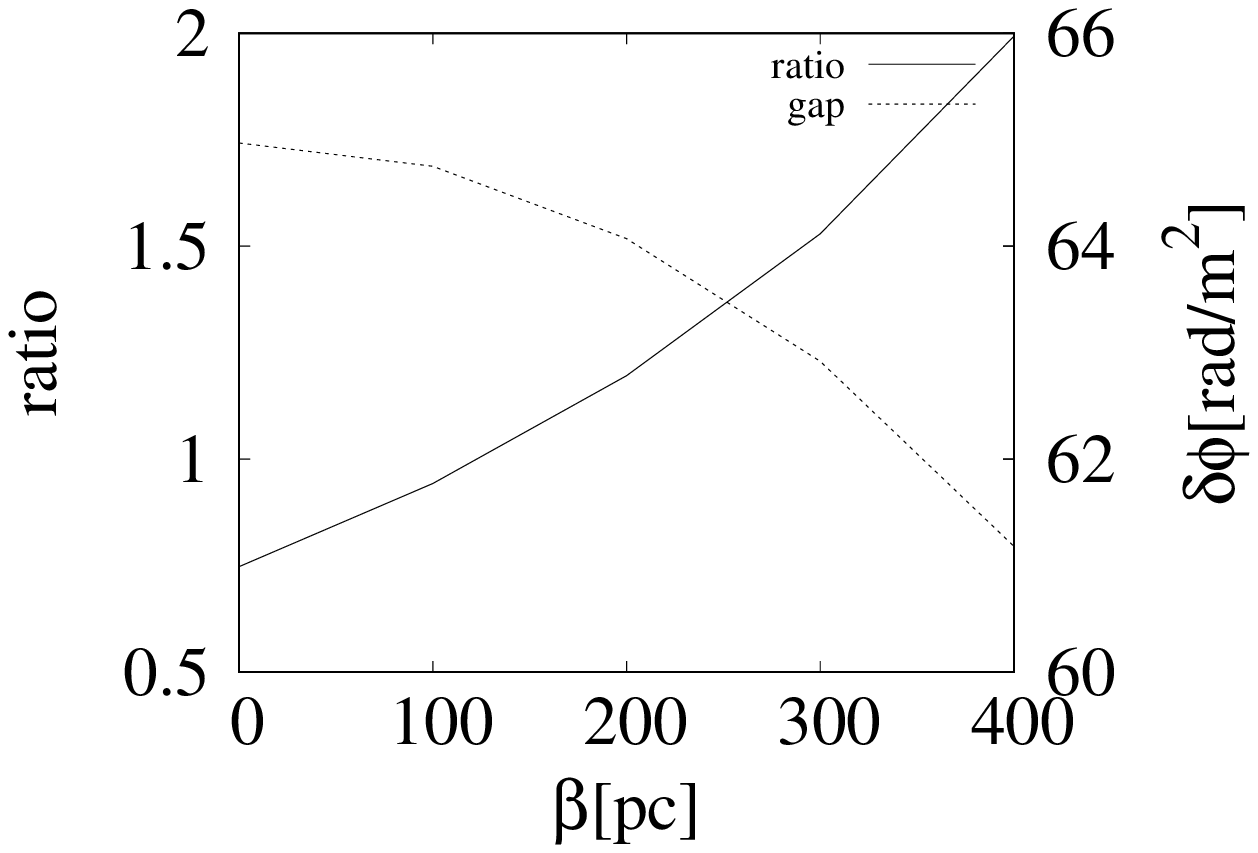}
 \caption{Same as Fig. \ref{fig:FDF_inc} but for varying the offset $\beta$. Here, we fix the inclination angle $\theta$ to 40~deg and $y =$ 200~pc.}
 \label{fig:FDF_offset-beta}
\end{figure*}

Finally, let us see the variation of the FDF when the offset of the line of sight from the galactic center is changed. Fig. \ref{fig:FDF_offset-y} shows the absolute value of the FDF (left panel) and the gap and peak-height ratio (right panel) for varying the offset $y$, where $\theta =$ 40~deg and $\beta =$ 0~pc are fixed. The distribution of $B_\parallel$ and Faraday depth along the line of sight are shown in Fig. \ref{fig:distribution_offset}.
The offset does not affect the maximum value of $B_{||}$ but its line of sight distribution becomes smoother and the overall value of $B_{||}$ becomes larger as $y$ increases. As a result, the gap increases significantly as 40, 65 and 90 [rad/m$^2$] for the offset of 100, 200 and 400 [pc], respectively. On the other hand, the heights of both peaks become smaller for a larger offset because a smaller number of layers contribute to the peaks. The change in the peak ratio is moderate ($\sim 20\%$). Fig. \ref{fig:FDF_offset-beta} shows the variation of the FDF where $\beta$ is shifted to 100, 200 and 400 pc with fixed values of $y$ = 200 pc and theta = 40 deg. Because the line of sight is inclined to $x$ direction (line of sight is $x$-$z$ plane parallel to $y$ direction), the FDF changes more sensitively with $y$ than $\beta$ (see Eq.~(\ref{eq:parallel})). Another reason why the variation in Fig. \ref{fig:FDF_offset-beta} is small compared to Fig. \ref{fig:FDF_offset-y} is that the $y$ value is fixed to 200 pc in Fig. \ref{fig:FDF_offset-beta} so that the distance from the center does not change largely by varing $\beta$ as 100, 200 and 400 pc.



\subsection{variant models}

In the previous sebsection, a simple model of a galaxy was adopted to understand the behavior of the FDF easily. We extend the fiducial model in several ways to consider more variations of observed galaxies.

\subsubsection{turbulent magnetic fields}

\begin{figure*}[t]
 \includegraphics[width=8cm]{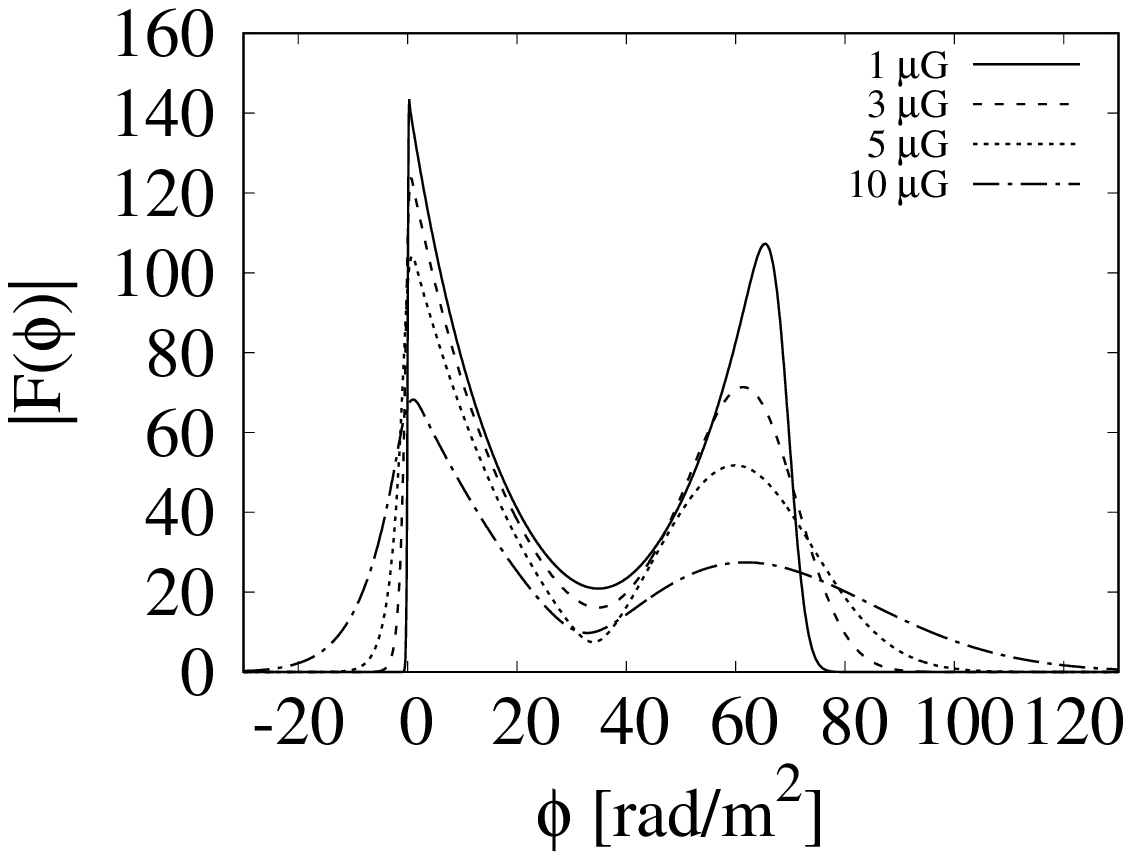}
 \includegraphics[width=8cm]{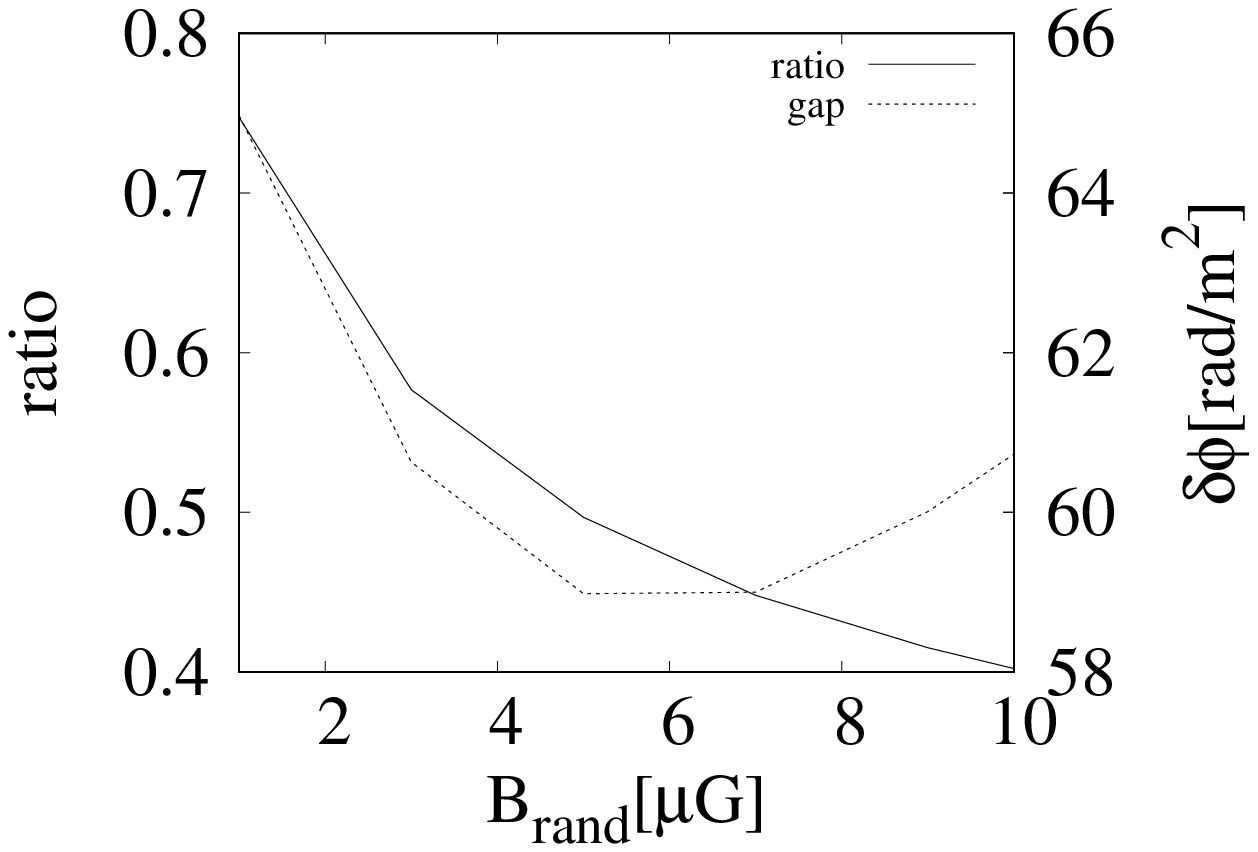}
 \caption{Same as Fig. \ref{fig:FDF_inc} but for $B_{\rm rand} =$ 1, 3, 5, and $10~\mu$G. The inclination angle $\theta$ is fixed to 40 deg.}
 \label{ran_absf}
\end{figure*}

In most galaxies, turbulent magnetic fields are comparable or stronger than coherent fields. In this subsection, we investigate the FDFs varying the amplitude of turbulent fields and fixing that of coherent fields. Fig. \ref{ran_absf} shows the absolute value of the FDFs with LOS component of turbulent magnetic fields, $B_{\rm rand}$, of 1, 3, 5 and $10~\mu {\rm G}$. For stronger turbulent fields, the width of the Gaussian function of each layer increases so that the peaks are widened and their heights are reduced. The peak widening is more significant for the right peak because it is contributed mostly from the far side of the galaxy. Hence, the ratio of the peak heights decreases with the increase of turbulent fields from 0.7 to 0.4 as $B_{\rm rand}$ changes from 0.1 $\mu$G to 10 $\mu$G. On the other hand, the position of Gaussian function of each layer is unchanged because the magnitude of coherent fields is fixed, so that the position of peaks do not change. Especially, the change in the gap is less than $10\%$ (see the range of the right-hand-side axis of the right panel). Thus, these FDFs share most of the features of the FDF of the fiducial model and the discussion given in the previous subsection applies to the current case as well.

\subsubsection{pitch angle}

\begin{figure}[t]
 \includegraphics[width=8cm]{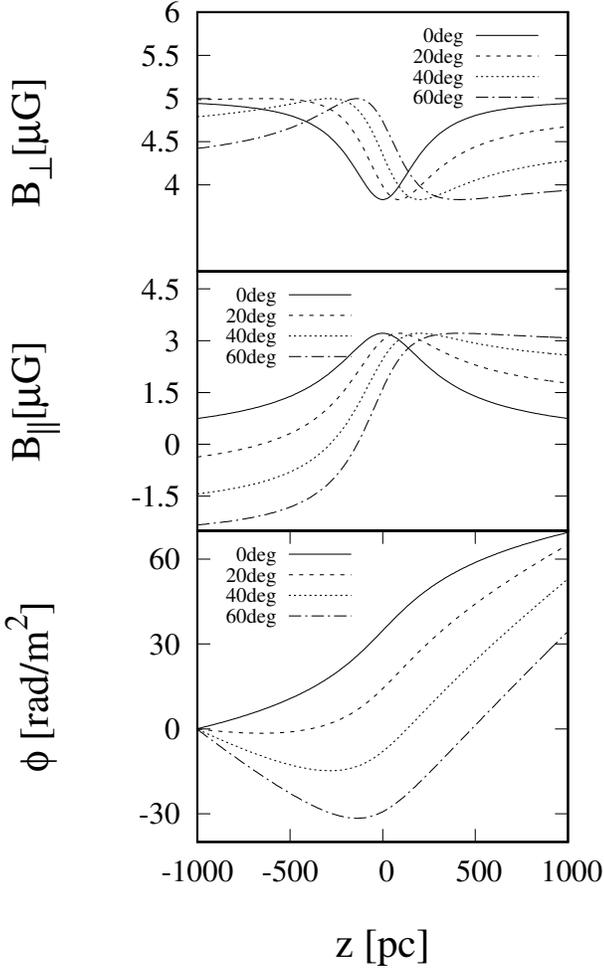}
 \caption{Same as Fig. \ref{fig:distribution} but with pitch angle $\theta'$ =  0, 20, 40 and 60 deg and fixed inclination $\theta = 40~{\rm deg}$. }
 \label{physics4}
\end{figure}

So far, we have considered the ring magnetic field model, but a lot of observations have implied that galactic magnetic fields have a variety of pitch angle, which can be realized by a rotation of the ring field at each position by the same amount \citep{Seigar,Eck,Chamandy2016,Chamandy}. \citet{Eck} and \citet{Chamandy2016} showed that the magnetic pitch angle of tens of galaxies is in the range of 10 $\sim$ 50 deg. In this section, we consider coherent magnetic fields with the pitch angle and simulate FDF changing the pitch angle systematically. Fig. \ref{physics4} shows the line-of sight distribution of perpendicular and parallel components of magnetic fields and $\phi(z)$ for pitch angle $\theta' = 0, 20, 40$ and $60~{\rm deg}$ with fixed inclination $\theta = 40~{\rm deg}$. It is seen that, for non zero pitch angle, the distribution of magnetic fields is not symmetric with respect to the galactic plane ($z = 0$) as opposed to the case with ring fields. The parallel component starts from a negative value and becomes positive at somewhere in the near side ($z < 0~{\rm pc}$). The transition point is closer to the galactic plane for a larger pitch angle. Consequently, Faraday depth first decreases from zero, takes the minimum value at the transition point and becomes positive in the end.

\begin{figure*}[t]
 \includegraphics[width=8cm]{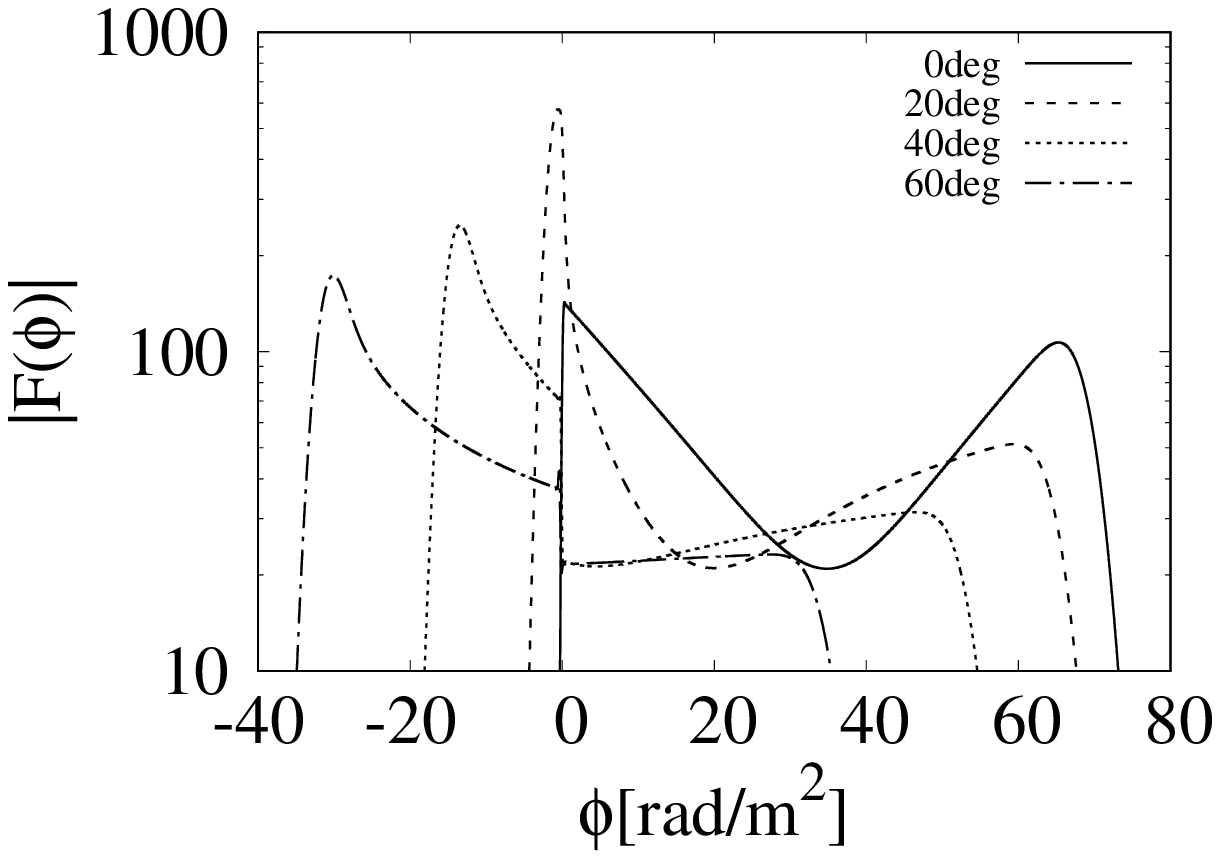}
 \includegraphics[width=8cm]{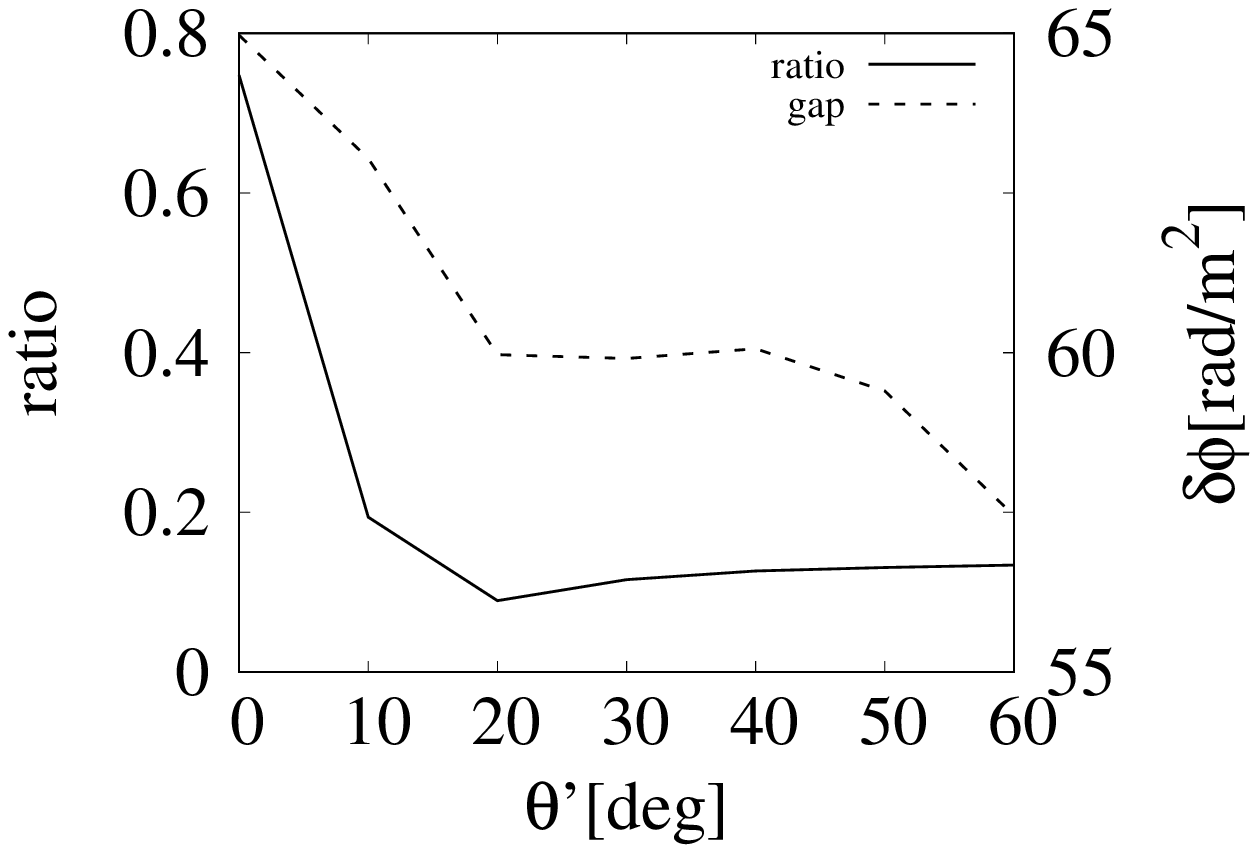}
 \caption{Same as Fig. \ref{fig:FDF_inc} but varying pitch angle $\theta'$ with fixed inclination $\theta = 40~{\rm deg}$.}
 \label{fdftheta40}
\end{figure*}

Resultant FDFs are shown in Fig.~\ref{fdftheta40}. Because Faraday depth becomes negative for non zero pitch angle, the left peak moves to negative $\phi$ region and is located at the minimum $\phi$. Therefore, the Faraday depth of the peak is smaller (larger $|\phi|$) for a larger pitch angle. The shift of the left peak from $\phi = 0$ reaches as large as 30 [rad/m$^2$] for $\theta' = 60$ deg. The Faraday depth of the right peak also moves leftward with the same amount. Hence, as we see in the right panel of Fig.~\ref{fdftheta40} the variation of the gap is very small and less than $10\%$ for 0 deg $\leq \theta' \leq$ 60 deg. On the other hand, the right peak has a lower height and is located at a smaller $\phi$ for a larger pitch angle. This is because only a narrow range of $z$ contributes to the right peak for a large pitch angle (see the bottom panel of Fig. \ref{physics4}). The magnitude of perpendicular component shown in the top panel of Fig. \ref{physics4} also affects the peak height, but this factor is very minor. It should be noted that the vertical axis of Fig. \ref{fdftheta40} is in logarithmic scale, while that of Fig. \ref{fig:distribution} is in linear scale. 

\begin{figure}[t]
 \includegraphics[width=8cm]{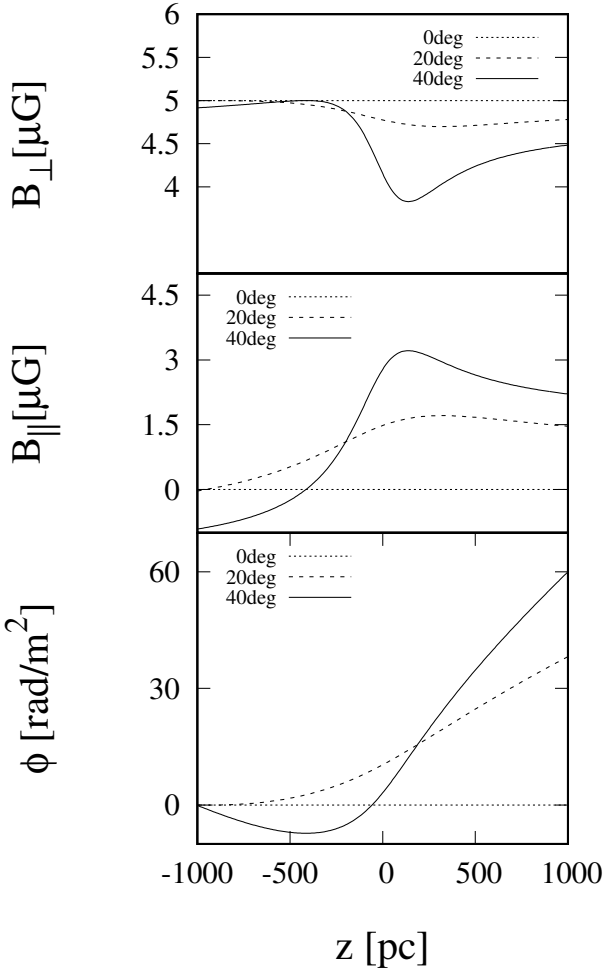}
 \caption{Same as Fig. \ref{fig:distribution} but with inclination $\theta$ = 0, 20 and 40 deg and fixed pitch angle $\theta' = 30~{\rm deg}$.}
 \label{phisics3}
\end{figure}

\begin{figure*}[t]
 \includegraphics[width=8cm]{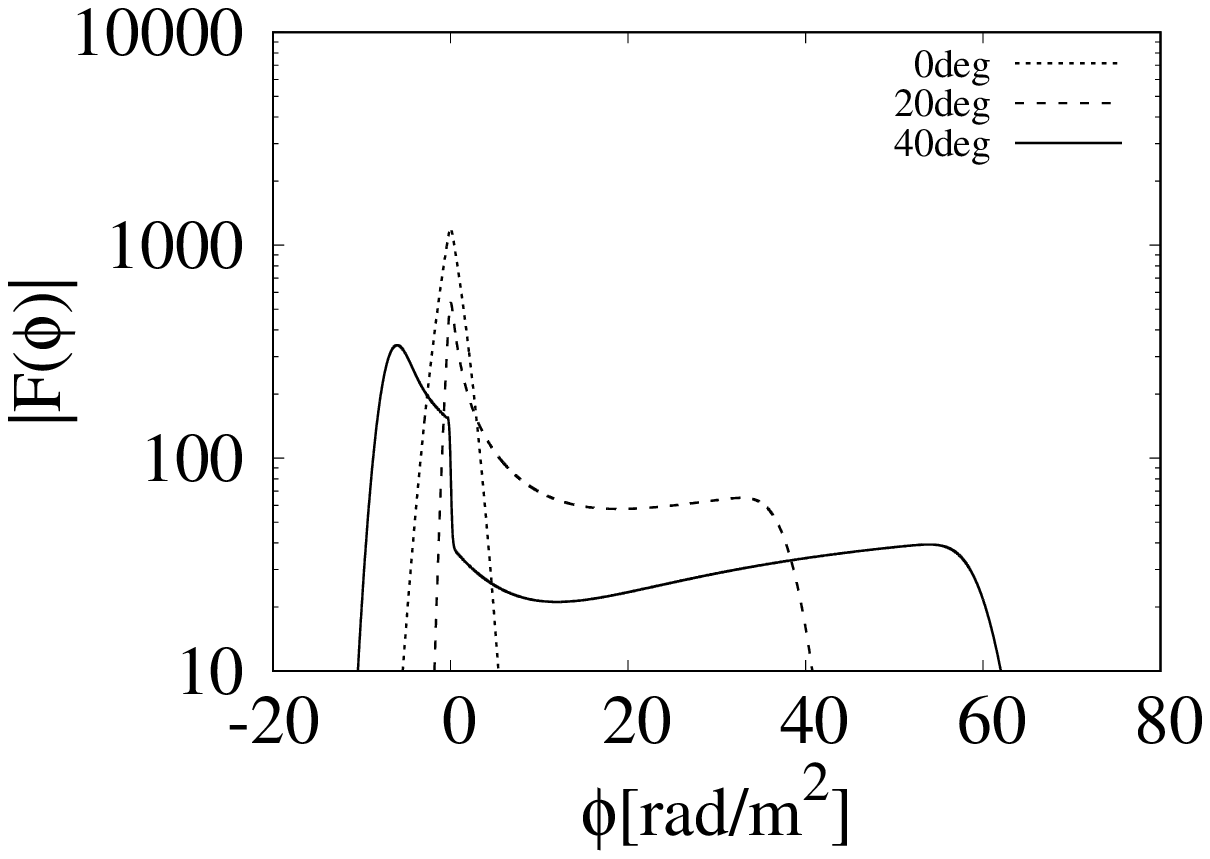}
 \includegraphics[width=8cm]{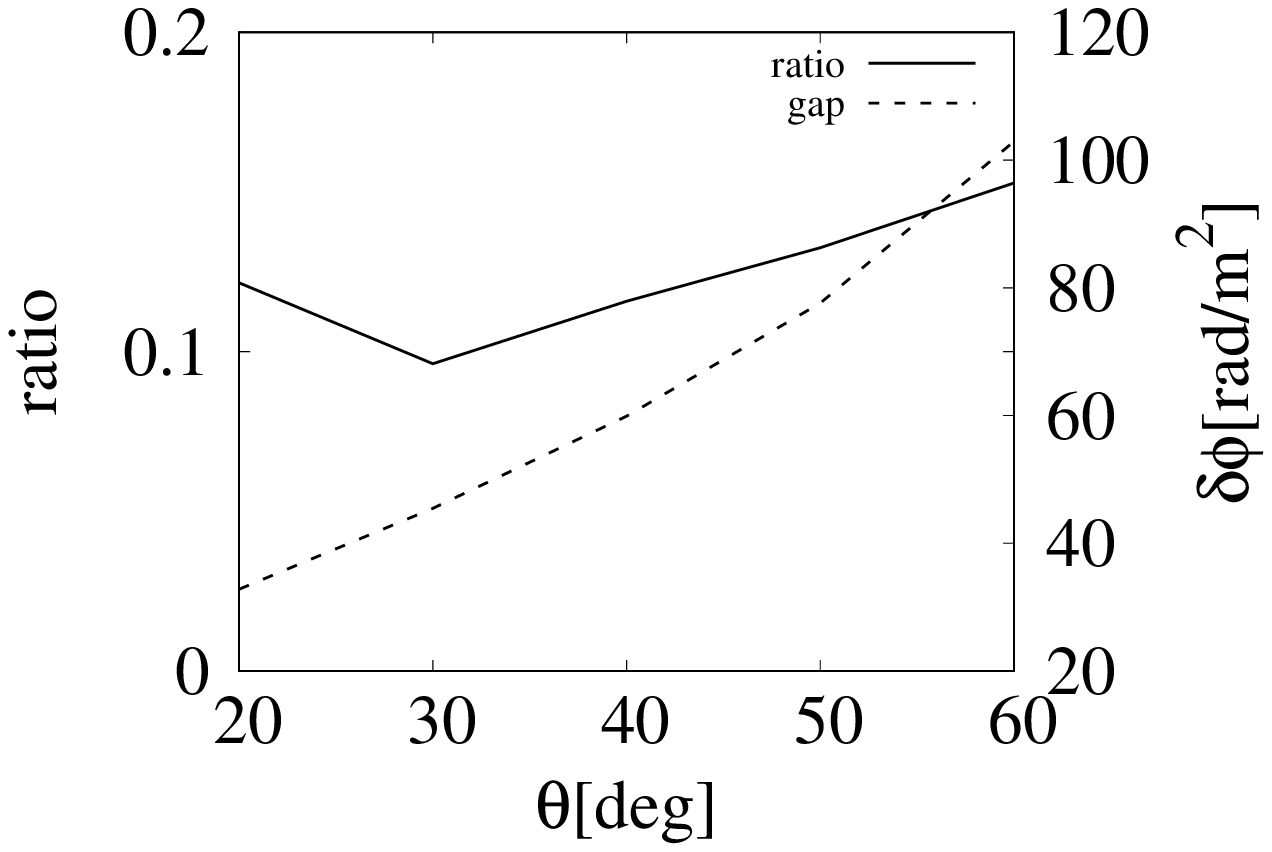}
 \caption{Same as Fig. \ref{fig:FDF_inc} but for varying inclination $\theta$ with fixed pitch angle $\theta' = 30~{\rm deg}$. Note that only one peak exists for $\theta \leq$ 10~deg.}
 \label{fdf30}
\end{figure*}

Figs. \ref{phisics3} and \ref{fdf30} are the comparison between different inclination angles ($\theta$ = 0, 20 and 40 deg) for a fixed value of the pitch angle of $\theta' = 30~{\rm deg}$. We can see that the effects of non zero pitch angle such as asymmetry of magnetic fields with respect to the galactic plane are suppressed for a smaller inclination angle. For $\theta$ = 0 (face-on), $B_{||} = 0$ and $B_{\perp}$ is constant so that the FDF reduces to the case of \cite{Ideguchi2017}, which has a single peak. The value of $B_{\perp}$ for $z > 0$ is significantly different between the three cases. For example, it is larger for $\theta =$ 20 deg than $\theta =$ 40 deg by about $15\%$, which leads to the difference in the right-peak heights by about $30\%$. This is reasonable considering the synchrotron emissivity is assumed to be proportional to $B_{\perp}^{1.8}$. The left-peak heights are also different by about the same factor due to the spread of the peak for $\theta =$ 40 deg. As a result, the variation of the peak ratio (right panel of Fig.~\ref{fdf30}) is less than $\sim 10\%$ as a function of $\theta$. A larger value of $B_{||}$ leads to a wider gap of the two peaks, which increases from 40 [rad/m$^2$] to 100 [rad/m$^2$] for $\theta =$ 20 deg and 60 deg.


\section{Summary and discussion}

In this paper, we investigated Faraday dispersion function (FDF) of disk galaxies extending a simple analytic model of galactic magnetic fields developed in \citet{Ideguchi2017}, which has axisymmetric coherent fields and turbulent fields. We studied the effects of inclination, relative amplitude of coherent and turbulent fields and pitch angle of coherent fields. Our model was rather simple but allowed us to understand the behavior of FDFs. First, we found FDFs have two peaks when galaxies are observed with non zero inclination. The gap and relative height of two peaks are dependent on the inclination angle and pitch angle. The relative amplitude of coherent and turbulent magnetic fields does not affect the qualitative features so much and widens the peaks.

It is rather surprising that FDFs exhibit a wide variety of shape just by varying the inclination and pitch angles. Especially, the fact that non zero inclination leads to two peaks in the FDF gives us an important caution that the number of peaks does not necessarily match to that of polarized sources. As a similar example, \citet{Beck2012a} reported that the FDF of a single polarized source can have two peaks in the presence of turbulent magnetic fields.

On the other hand, it is well known that reconstruction of the FDF from observed polarization spectrum can produce artificial multiple peaks due to the incomplete coverage of wavelength range, even if the intrinsic FDF has a single peak \citep{Brentjens2005}. Especially, this happens when the width of the intrinsic FDF is larger than the full width at half maximum of RM spread function determined by the wavelength coverage. Simulations of reconstruction of more complicated FDFs found here will be given elsewhere.

In this work, we ignored galactic halos (or thick disks). From observations of edge-on galaxies, distribution of emission vertical to the galactic disk is described by exponential functions, and the emission at 2 kpc from the disk, for instance, is $\sim$ 10 times smaller than that from the disk \citep{Dumke1998,Heesen2009,Irwin2012a,Irwin2013,Wiegert2015,Krause2018}. This suggests that the contribution of halo emission is substantially small and thus, it barely affects to the results of this work. On the other hand, contribution of the magnetic field in the halos is not trivial. For instance, \citet{Fletcher2011} suggested that the magnetic field topology of disk and halo are different by the observation of nearby galaxy M51. If the LOS component of halo magnetic field is coherent and polarized emission from the halo can be ignored, the existence of the halo causes just the shift of the FDF in the Faraday depth space without changing the shape. On the other hand, if it is turbulent, this would lead to Faraday dispersion, which makes the FDF broader and reduces its the absolute value, and consequently the shape of FDF would become more complicated.

Another simple assumption we made in this work is that the physical quantities other than the turbulent magnetic fields are uniform. Regarding the synchrotron polarization emissivity, the observations of edge-on galaxies show the exponential decay from the disk \citep{Dumke1998,Heesen2009,Irwin2012a,Irwin2013,Wiegert2015,Krause2018}. As we saw in the previous section, the near-side and far-side of the galaxy contribute to the left and right peaks, respectively. Thus, the smaller emissivity in the off-disk regions reduces the amplitude of the two peaks. The thermal electron density ($n_{\rm e}$) is also though to decay from the disk by observations of the Milky Way \citep[e.g.][]{Gaensler2008}. The decay of the density from the disk results in the Faraday depth does not accumulate in the off-disk regions and thus, the width of the two peaks can get smaller. On the other hand, if $n_{\rm e}$ is turbulent, this affects the Faraday dispersion in the same way as the turbulent component of LOS magnetic fields. This effect would also be statistically averaged out since we added up the FDFs from $N_\perp \times N_\perp = 100$ cells in the plane of the sky.

Combining Faraday tomography and 2-dimensional imaging, we will be able to obtain Faraday cube, that is, polarization emissivity as a function of Faraday depth and position in the sky. This will allow us probe 3-dimensional structure of galactic magnetic fields, such as topology of global fields and the position dependence of the strength of turbulent fields. Investigation of Faraday cube is also part of our future work.

\section*{Acknowledgement}
KT is partially supported by JSPS KAKENHI Grant Numbers JP15H05896, JP16H05999 and JP17H01110, and Bilateral Joint Research Projects of JSPS. YM  is supported in part by Grand-in-Aid from JSPS Research Fellow, No. 17J06936.


\end{document}